\documentstyle[12pt]{article}
 
\begin{document} 

\baselineskip 22pt 

\begin{center}
{\Large
\bf Solving the Puzzle of
${M_{D^*}-M_{D}\over M_{D_s^*}-M_{D_s}}$ $\simeq$
${M_{B^*}-M_{B}\over M_{B_s^*}-M_{B_s}}$ $\simeq$ $1$}\\
\vspace{1.0cm}
Dae Sung Hwang$^*$ and Gwang-Hee Kim$^{\dagger}$\\
{\it{Department of Physics, Sejong University, Seoul 143--747,
Korea}}\\
\vspace{2.0cm}
{\bf Abstract}\\
\end{center}

The commonly used Hamiltonian of the chromomagnetic hyperfine
splitting is inversely proportional to the product of the
masses of two constituent quarks composing the meson.
So it is expected to have
$(M_{D^*}-M_{D})/(M_{D_s^*}-M_{D_s})$ $\simeq$
$(M_{B^*}-M_{B})/(M_{B_s^*}-M_{B_s})$ $\simeq$ $1.6$,
when the constituent quark masses
$m_{u,d}=0.33$ GeV and $m_s=0.53$ GeV are used.
However, the experimental results show that the above ratios are
very close to 1.
We solve this puzzle by employing the Hamiltonian
recently proposed by Scora and Isgur. 
\\

\vfill 

\noindent
PACS number(s): 12.39.Ki, 12.40.Yx, 14.40.Lb, 14.40.Nd

\vspace*{0.5cm}

\noindent
$^*$e-mail: dshwang@phy.sejong.ac.kr\\
$^{\dagger}$e-mail: gkim@phy.sejong.ac.kr
\thispagestyle{empty} 
\pagebreak 

\baselineskip 22pt

Review of Particle Physics \cite{rpp}
presents the masses of the $D$ mesons (MeV):
\begin{eqnarray}
& &M_{D^\pm}=1869.3\pm 0.5\, ,\ \
M_{D^0}=1864.5\pm 0.5\, ,\ \
M_{D_s^\pm}=1968.5\pm 0.6\, ,
\nonumber\\
& &M_{D^{*\pm}}=2010.0\pm 0.5\, ,\ \
M_{D^{*0}}=2006.7\pm 0.5\, ,\ \
M_{D_s^{*\pm}}=2112.4\pm 0.7\, ,
\label{p1}\\
& &M_{D^{*\pm}}-M_{D^\pm}=140.64\pm 0.09\, ,\ \
M_{D^{*0}}-M_{D^0}=142.12\pm 0.07\, ,
\nonumber\\
& &M_{D_s^{*\pm}}-M_{D_s^\pm}=143.8\pm 0.4\, ,
\nonumber
\end{eqnarray}
and the masses of the $B$ mesons (MeV):
\begin{eqnarray}
& &M_{B^\pm}=5278.9\pm 1.8\, ,\ \
M_{B^0}=5279.2\pm 1.8\, ,\ \
M_{B_s^0}=5369.3\pm 2.0\, ,
\nonumber\\
& &M_{B^{*}}=5324.8\pm 1.8\, ,\ \
M_{B_s^{*0}}=5416.3\pm 3.3\, ,
\label{p2}\\
& &M_{B^{*}}-M_{B}=45.7\pm 0.4\, ,\ \
M_{B_s^{*0}}-M_{B_s^0}=47.0\pm 2.6\, ,
\nonumber
\end{eqnarray}
where $M_{B}\equiv (M_{B^\pm}+M_{B^0})/2\, $.
We note that $M_{D_s^{*\pm}}-M_{D_s^\pm}$ is very close to
$M_{D^{*\pm}}-M_{D^\pm}$ and $M_{D^{*0}}-M_{D^0}$ in (\ref{p1}),
and that $M_{B_s^{*0}}-M_{B_s^0}$ to $M_{B^{*}}-M_{B}$ in (\ref{p2}).

The commonly used Hamiltonian for the chromomagnetic hyperfine
splitting of vector and pseudoscalar mesons is given by
\cite{rujula}
\begin{equation}
H_{hf}({\bf r})={32\pi\alpha_s\over 9m_Qm_{\bar{q}}}\,
{\bf s}_Q\cdot {\bf s}_{\bar{q}}\,
\delta ({\bf r}),
\label{p3}
\end{equation}
where $\alpha_s={\bar{g}}^2/4\pi$ of the QCD running coupling constant
${\bar{g}}$, and $m_Q\ ({\bf s}_Q)$ and $m_{\bar{q}}\ ({\bf s}_{\bar{q}})$
are the masses (spins) of the heavy quark $Q$ and the light antiquark
${\bar{q}}$ inside a heavy-light meson, respectively.
The Hamiltonian in (\ref{p3}) corresponds to the potential due to
one-gluon exchange for s-wave bound states.
Treating the chromomagnetic hyperfine splitting as a perturbation,
one obtains
\begin{equation}
M_{M^*}-M_M={32\pi\alpha_s\over 9m_Qm_{\bar{q}}}\,
|\Psi_{Q{\bar{q}}}({\bf 0})|^2,
\label{p4}
\end{equation}
where $\Psi_{Q{\bar{q}}}({\bf 0})$ is the two-body bound state wave function
at origin.
When it is assumed that
$D_{u,\, d}$ and $D_s$ (or $B_{u,\, d}$ and $B_s$) have similar values of
$\alpha_s$ and $|\Psi_{Q{\bar{q}}}({\bf 0})|$,
$M_{M^*}-M_M$ is proprtional to $1/m_{\bar{q}}$
(${\bar{q}}={\bar{u}},\ {\bar{d}},\ {\bar{s}}$)
for $D$ (or $B$) mesons.
Then it is expected to have
$(M_{D^*}-M_{D})/(M_{D_s^*}-M_{D_s})$ $\simeq$
$(M_{B^*}-M_{B})/(M_{B_s^*}-M_{B_s})$ $\simeq$ $1.6$,
when the constituent quark masses
$m_{u,d}=0.33$ GeV and $m_s=0.53$ GeV are used.
However, the experimental results in (\ref{p1}) and (\ref{p2}) give
the values
\begin{equation}
{M_{D^{*\pm}}-M_{D^\pm}\over M_{D_s^{*\pm}}-M_{D_s^\pm}}
=0.978\pm 0.003,\qquad
{M_{B^{*}}-M_{B}\over M_{B_s^{*0}}-M_{B_s^0}}
=0.972\pm 0.054,
\label{p4a}
\end{equation}
which are very close to 1.
Goity and Hou \cite{goity}
noticed this peculiar property of the heavy meson masses.
Randall and Sather \cite{randall}
emphasized this discrepancy and called it a puzzle.
They suggested that these data might be an interesting probe of heavy
mesons.
The purpose of this letter is to
solve this puzzle by employing the Hamiltonian
recently proposed by Scora and Isgur.

Scora and Isgur \cite{scora}
proposed the following ${\bar{H}}_{hf}({\bf r})$ as a modification
of $H_{hf}({\bf r})$ in (\ref{p3}):
\begin{equation}
{\bar{H}}_{hf}({\bf r})=
[{m_Qm_{\bar{q}}\over E_QE_{\bar{q}}}]^{1\over 2}
\Bigl( {32\pi a\alpha_s\over 9m_Qm_{\bar{q}}}\,
{\bf s}_Q\cdot {\bf s}_{\bar{q}}\,
\delta ({\bf r})\Bigr)
[{m_Qm_{\bar{q}}\over E_QE_{\bar{q}}}]^{1\over 2},
\label{p5}
\end{equation}
where the term in the parentheses would be the ordinary Fermi contact
term in (\ref{p3}) if the anomalous coupling coefficient $a$ were unity,
and where $E_i=(m_i^2+{\bf p}^2)^{1\over 2}$.
For s-wave bound states, the expectation value of the Hamiltonian
${\bar{H}}_{hf}({\bf r})$ in (\ref{p5}) is given by \cite{scora}
\begin{equation}
\langle {\bar{H}}_{hf}({\bf r})\rangle =
[{2S(S+1)-3\over 4}]({32\pi a\alpha_s\over 9m_Qm_{\bar{q}}})
|{\bar{\Psi}}_{Q{\bar{q}}}({\bf 0})|^2,
\label{p6}
\end{equation}
\begin{equation}
{\bar{\Psi}}_{Q{\bar{q}}}({\bf 0})={1\over (2\pi )^{3/2}}
\int d^3{\bf p}\, [{m_Qm_{\bar{q}}\over E_QE_{\bar{q}}}]^{1\over 2}\,
\Phi_{Q{\bar{q}}}({\bf p}),
\label{p7}
\end{equation}
where $S=0$ or 1 is the total spin of the meson.

In order to obtain the heavy meson wave function
$\Phi_{Q{\bar{q}}}({\bf p})$ in (\ref{p7}),
Scora and Isgur \cite{scora}
applied the variational method to the Hamiltonian with
the nonrelativistic kinetic energy terms.
They adopted the Gaussian wave function as a trial wave function
for the heavy meson ground state.
We follow the same procedure as theirs in obtaining the heavy meson
ground state wave function, except for that we use the relativistic
kinetic energy terms since the velocity of the light quark inside heavy
meson is large.
We apply the variational
method to the relativistic Hamiltonian
\begin{equation}
H={\sqrt{{\bf p}^2+{m_Q}^2}}+{\sqrt{{\bf p}^2+{m_{\bar{q}}}^2}}+V(r),
\label{g13}
\end{equation}
where ${\bf r}$ and ${\bf p}$ are the relative coordinate and its
conjugate momentum.
We take the following potential energy for $V(r)$ in (\ref{g13}),
\begin{equation}
V(r)=-{\alpha_c\over r}+Kr+{\bar{H}}_{hf}(r),
\label{g13a}
\end{equation}
where ${\bar{H}}_{hf}(r)$ is the modified chromomagnetic hyperfine
Hamiltonian in (\ref{p5}) proposed by Scora and Isgur.
We take the Gaussian wave function as a variational wave function
for the heavy meson ground state,
\begin{equation}
\Psi ({\bf r})=({{\beta}\over {\sqrt{\pi}}})^{3/2}e^{-{\beta}^2r^2/2},
\qquad
\Phi ({\bf{p}})={1\over ( \sqrt{\pi} \beta )^{3/2}}
e^{-{\bf{p}}^2/2{\beta}^2}.
\label{f2aa}
\end{equation}
The ground state wave function is then given by \cite{scora,hknhk}
\begin{equation}
\langle H\rangle =\langle\psi\vert H\vert\psi\rangle =E(\beta ),
\qquad
{dE(\beta )\over d\beta} =0 \ {\rm at}\ \beta ={\bar{\beta}},
\label{f2aaa}
\end{equation}
where ${\bar{\beta}}$
represents the inverse size of the meson
($\langle r^2 {\rangle}^{1/2}=3/(2\, {\bar{\beta}})$),
and $\bar E \equiv E({\bar{\beta}})$ the meson mass.
In explicit calculations, we used the following
four different potential models
which have the values of $\alpha_c$, $K$ and quark masses given in
Table 1:
(A) Scora and Isgur \cite{scora},
(B) Eichten $et$ $al.$ \cite{eich},
(C) Hagiwara $et$ $al.$ \cite{hagi},
and
(D) Model D which is the same as the Hagiwara $et$ $al.$'s model (C)
except for the values $\alpha_c(D)=0.48$ and $\alpha_c(B)=0.32$
which are given from the running coupling constants at the energy
scales of $M_D$ and $M_B$, respectively.

For each of the four models of (A), (B), (C), and (D),
we obtained the function $E(\beta )$ in (\ref{f2aaa}).
As a representative, we present $E(\beta )\equiv M$ for the model (C) of
Hagiwara $et$ $al.$ in Fig 1, where (a), (b), (c), and (d)
correspond to $D_d$, $D_s$, $B_d$, and $B_s$ mesons, respectively.
In Fig 1, the graphs with the mark $\Box$ are what we obtain when
we take the anomalous coupling coefficient $a$ in (\ref{p5}) as zero,
that is, when we do not include the ${\bar{H}}_{hf}(r)$ term in
(\ref{g13a}).
The values of the variational parameter $\beta$ which give the minimum
of $E(\beta )$ are presented in the
$D_d(0)$, $D_s(0)$, $B_d(0)$, and $B_s(0)$ columns of Table 2.

The graphs with the marks $\Diamond$ and $+$ in Fig. 1 correspond
to those for the vector and pseudoscalar mesons, respectively, which
we obtain when we use desirable values of $a$, as we explain in the
following.
By including of the ${\bar{H}}_{hf}(r)$ term in (\ref{g13a})
with a fixed value of the anomalous coupling coefficient $a$ in (\ref{p5}),
we obtain $E(\beta )\equiv M$ and its minimum from the condition in
(\ref{f2aaa}),
for the vector and pseudoscalar mesons separately.
Then, by performing the same procedure with different values of $a$,
we obtain the vector and pseudoscalar meson masses $M_{M^*}$ and $M_M$
which are given by the minimum of $E(\beta )$,
and their difference $\Delta M\equiv M_{M^*}-M_M$ as functions of the
coefficient $a$.
We obtained these funtions for each of the four models of (A), (B), (C),
and (D). In Fig. 2, as a representative, we present
$\Delta M\equiv M_{M^*}-M_M$ as functions of $a$
for $D_d$, $D_s$, $B_d$, and $B_s$ mesons obtained for the model (C).

The nice feature of Fig. 2 is that the graph of $D_d$ is very close to
that of $D_s$, also for $B_d$ and $B_s$, and this interesting character
is the clue for
solving the puzzle in this letter.
The dotted vertical lines indicate the values of $a$ which give rise to
the experimental values of $\Delta M$.
In Table 3,
we present these values of $a$ and $\Delta M\equiv M_{M^*}-M_M$ obtained
by these $a$ values.
Table 3 shows that 
the ratio $R\equiv \Delta M(M_d)/\Delta M(M_s)$ are obtained as the values
which are
very close to 1 for both $D$ and $B$ mesons, and the puzzle is solved.
The graphs with the marks $\Diamond$ and $+$ in Fig. 1 are 
$E(\beta )\equiv M$ obtained
with the values of $a$ in Table 3.
Fig. 3 shows that the minimum of $E(\beta )$ for the vector meson
is bigger than that for the pseudoscalar meson as it should be,
and the value of the corresponding ${\bar{\beta}}$
(denoted by dotted vertical lines) for the pseudoscalar meson
is bigger than that for the vector meson.
This means that the size of the pseudoscalar meson is smaller than that
of the vector meson, 
since $\langle r^2 {\rangle}^{1/2} =3/(2\, {\bar{\beta}})$.
We present the sizes of the vector and pseudoscalar mesons and their
ratios in Table 4.

In conclusion, we have solved the puzzle of
$(M_{D^*}-M_{D})/(M_{D_s^*}-M_{D_s})$ $\simeq$
$(M_{B^*}-M_{B})/(M_{B_s^*}-M_{B_s})$ $\simeq$ $1$
by adopting the modified chromomagnetic hyperfine
Hamiltonian in (\ref{p5}) proposed by Scora and Isgur \cite{scora}
instead of the commonly used
chromomagnetic hyperfine Hamiltonian in (\ref{p3}).\\

\pagebreak
\vspace*{1.0cm}

\noindent
{\em Acknowledgements} \\
\indent
D.S. Hwang is grateful to Wei-Shu Hou for useful discussions.
This work was supported
in part by the Basic Science Research Institute Program,
Ministry of Education, Project No. BSRI-96-2414,
and in part by Non-Directed-Research-Fund,
Korea Research Foundation 1996.\\

\pagebreak

\pagebreak

\begin{table}[h]
\begin{center}
\begin{tabular}{|c|c|c|c|c|c|c|}   \hline
        &$\alpha_c$&$K$&$m_u=m_d$
        &$m_s$&$m_c$&$m_b$
\\   \hline
Scora    &0.67&0.18&330&550&1820&5200\\
Eichten  &0.52&0.18&330&530&1840&5180\\
Hagiwara &0.47&0.19&330&530&1320&4750\\
Model D  &($D$)0.48,\ ($B$)0.32&0.19&330&530&1320&4750\\
\hline
\end{tabular}
\end{center}
\caption{The values of the parameters in the potential
Eq. (\protect\ref{g13a}) of the models used in this letter.}
\end{table}

\begin{table}[h]
\vspace*{1.2cm}
\hspace*{-0.8cm}
\begin{tabular}{|c|c|c|c|c|c|c|c|c|c|c|c|c|}   \hline
        &$D_d(0)$&$D_d^*$&$D_d$
        &$D_s(0)$&$D_s^*$&$D_s$
        &$B_d(0)$&$B_d^*$&$B_d$
        &$B_s(0)$&$B_s^*$&$B_s$
\\   \hline
Scora    &556&526&715&592&561&767
         &663&646&729&716&697&794\\
Eichten  &515&488&662&547&518&709
         &591&577&647&636&620&701\\
Hagiwara &484&460&605&512&487&642
         &574&561&625&616&601&673\\
Model D  &486&462&608&514&490&639
         &523&511&567&558&545&608\\
\hline
\end{tabular}
\caption{The values of the variational parameter ${\bar{\beta}}$ (MeV)
of the Gaussian wave function which
minimize $\langle H\rangle$.
Here, the values in the column of $D_d(0)$ are ${\bar{\beta}}$'s
obtained without taking ${\bar{H}}_{hf}(r)$ into account,
and those in the columns of $D_d^*$ and $D_d$ are ${\bar{\beta}}$'s
obtained with taking ${\bar{H}}_{hf}(r)$ into account for vector and
pseudoscalar mesons, respectively,
and the same for other mesons.
}
\end{table}

\begin{table}[h]
\vspace*{0.8cm}
\hspace*{-1.5cm}
\begin{tabular}{|c|c|c|c|c|c|c|c|c|}   \hline
    &$a(D)$&$\Delta M(D_d)$&$\Delta M(D_s)$&$R(D)$
    &$a(B)$&$\Delta M(B_d)$&$\Delta M(B_s)$&$R(B)$
\\  \hline
Scora    &1.21&141&144&0.98&0.78&46.0&49.5&0.93\\
Eichten  &1.85&143&144&0.99&1.27&45.7&48.2&0.95\\
Hagiwara &1.86&147&143&1.03&1.37&45.3&47.2&0.96\\
Model D  &1.80&146&143&1.02&2.49&45.6&46.5&0.98\\ \hline
Expts.&---&$141\pm 0.1$&$144\pm 0.4$&0.98$\pm 0.00$
&---&$45.7\pm 0.4$&$47.0\pm 2.6$&0.97$\pm 0.05$\\
\hline
\end{tabular}
\caption{The obtained values of anomalous coupling constant $a$,
the mass differences
$\Delta M\equiv M_{M^*}-M_M$ (MeV) of the vector and
pseudoscalar mesons,
and the ratio $R\equiv \Delta M(M_d)/\Delta M(M_s)$.}
\end{table}

\pagebreak

\begin{table}[h]
\vspace*{1.2cm}
\hspace*{-0.7cm}
\begin{tabular}{|c|c|c|c|c|c|c|c|c|c|c|c|c|}   \hline
        &$D_d^*$&$D_d$&${D_d^*\over D_d}$
        &$D_s^*$&$D_s$&${D_s^*\over D_s}$
        &$B_d^*$&$B_d$&${B_d^*\over B_d}$
        &$B_s^*$&$B_s$&${B_s^*\over B_s}$
\\   \hline
Scora    &2.85&2.10&1.36&2.67&1.96&1.36
         &2.32&2.06&1.13&2.15&1.89&1.14\\
Eichten  &3.07&2.27&1.35&2.90&2.12&1.37
         &2.60&2.32&1.12&2.42&2.14&1.13\\
Hagiwara &3.26&2.48&1.31&3.08&2.34&1.32
         &2.67&2.40&1.11&2.50&2.23&1.12\\
Model D  &3.25&2.47&1.32&3.06&2.35&1.30
         &2.94&2.65&1.11&2.75&2.47&1.11\\ \hline
Average  &3.11&2.33&1.34&2.93&2.19&1.34
         &2.63&2.36&1.12&2.46&2.18&1.13\\
\hline
\end{tabular}
\caption{The sizes of mesons $\langle r^2 {\rangle}^{1/2}$
($=3/(2\, {\bar{\beta}})$)
in the unit of
${\rm GeV}^{-1}$ ($1\ {\rm GeV}^{-1}=0.197\ {\rm fm}$)
obtained from the values of ${\bar{\beta}}$'s
given in Table 2, and the ratios of the vector and
pseudoscalar meson sizes.
}
\end{table}

\pagebreak

\noindent
{\large\bf
Figure Captions}\\

\noindent
Fig. 1. The meson masses $E(\beta )\equiv M$ as functions of the
variational parameter $\beta$ for (a) $D_d$, (b) $D_s$, (c) $B_d$,
and (d) $B_s$ mesons.\\

\noindent
Fig. 2. The difference of the vector and pseudoscalar meson masses
$\Delta M\equiv M_{M^*}-M_M$ as functions of
the anomalous coupling coefficient $a$
for $D_d$, $D_s$, $B_d$, and $B_s$ mesons.\\

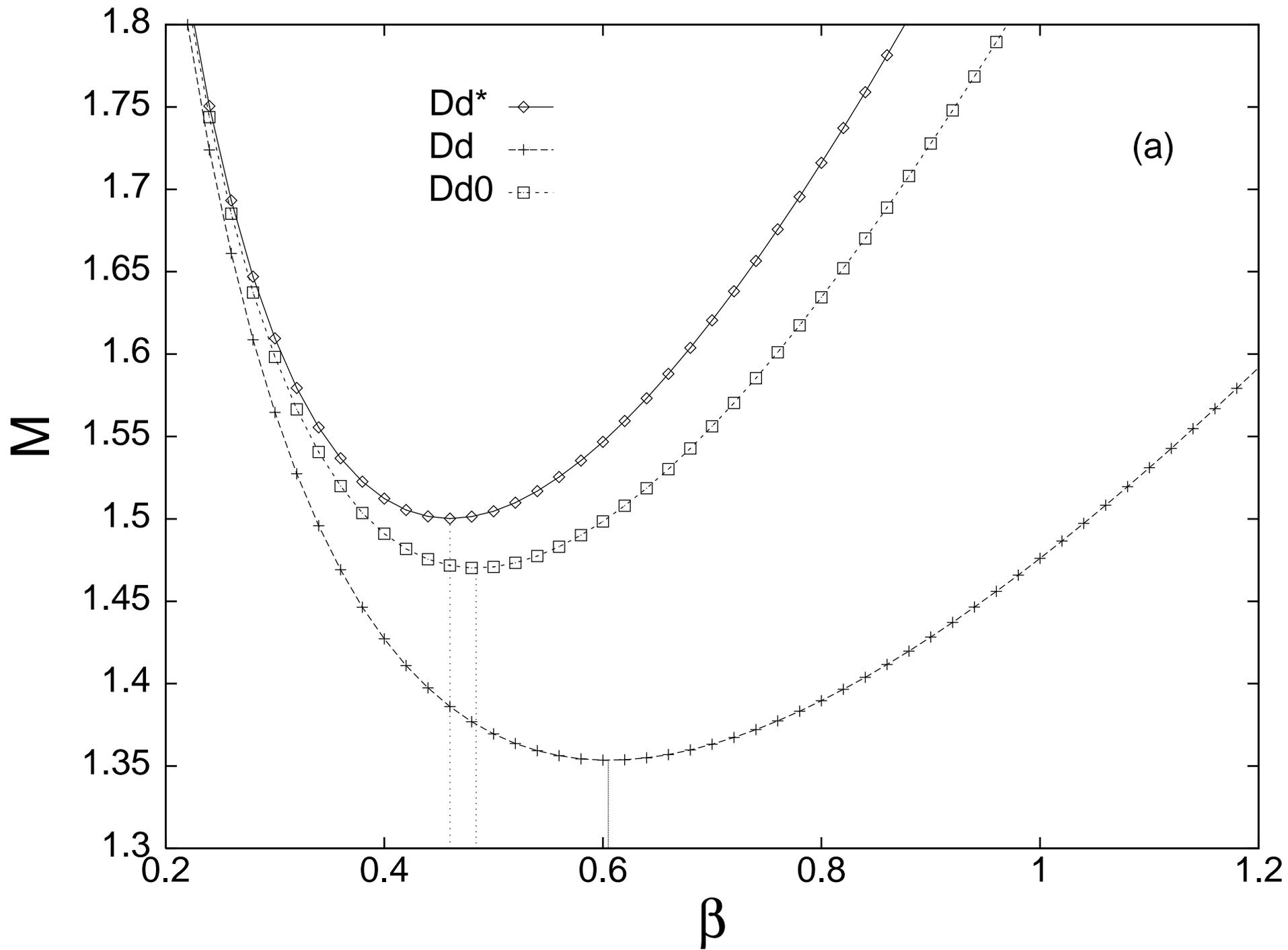
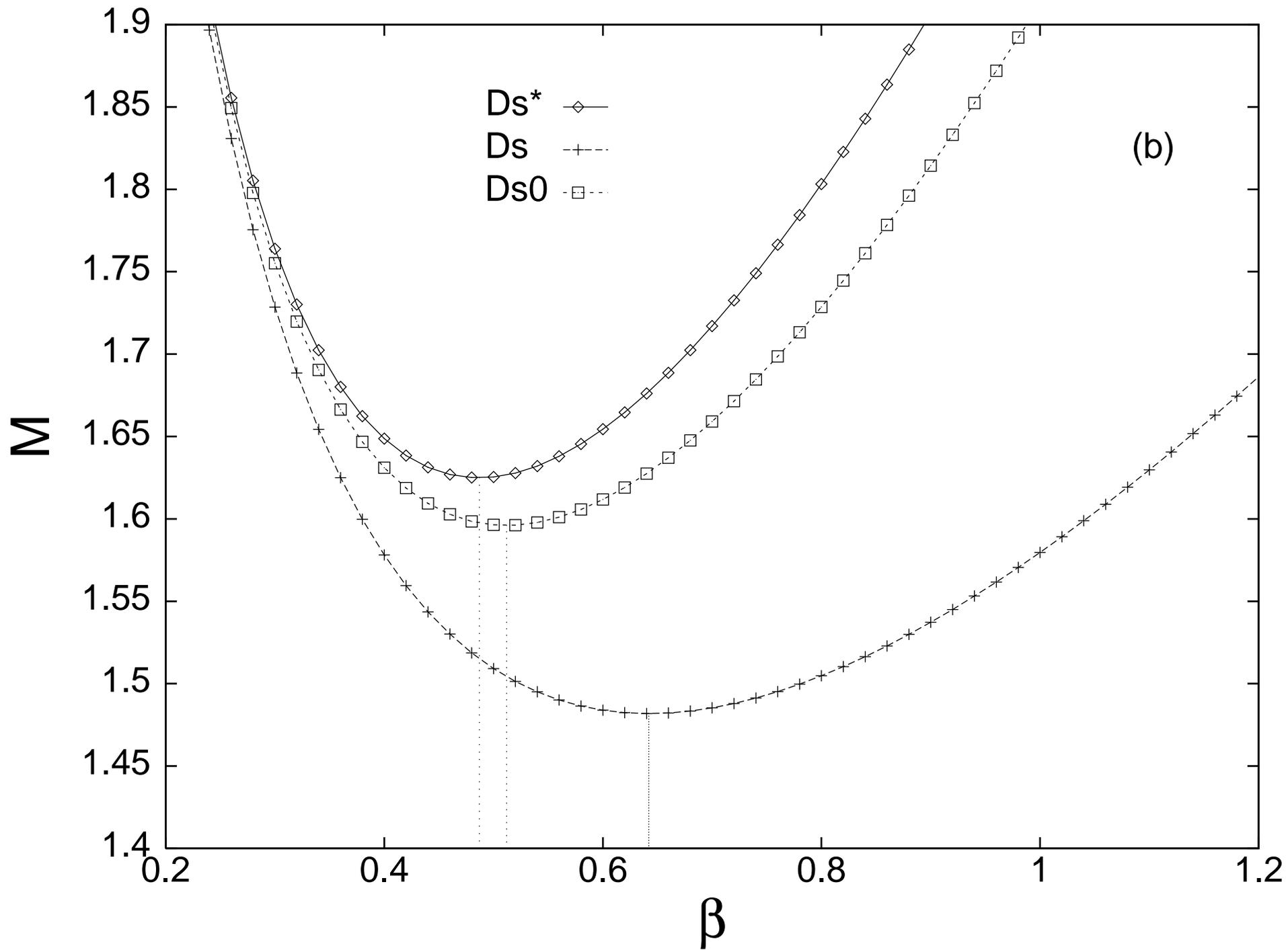
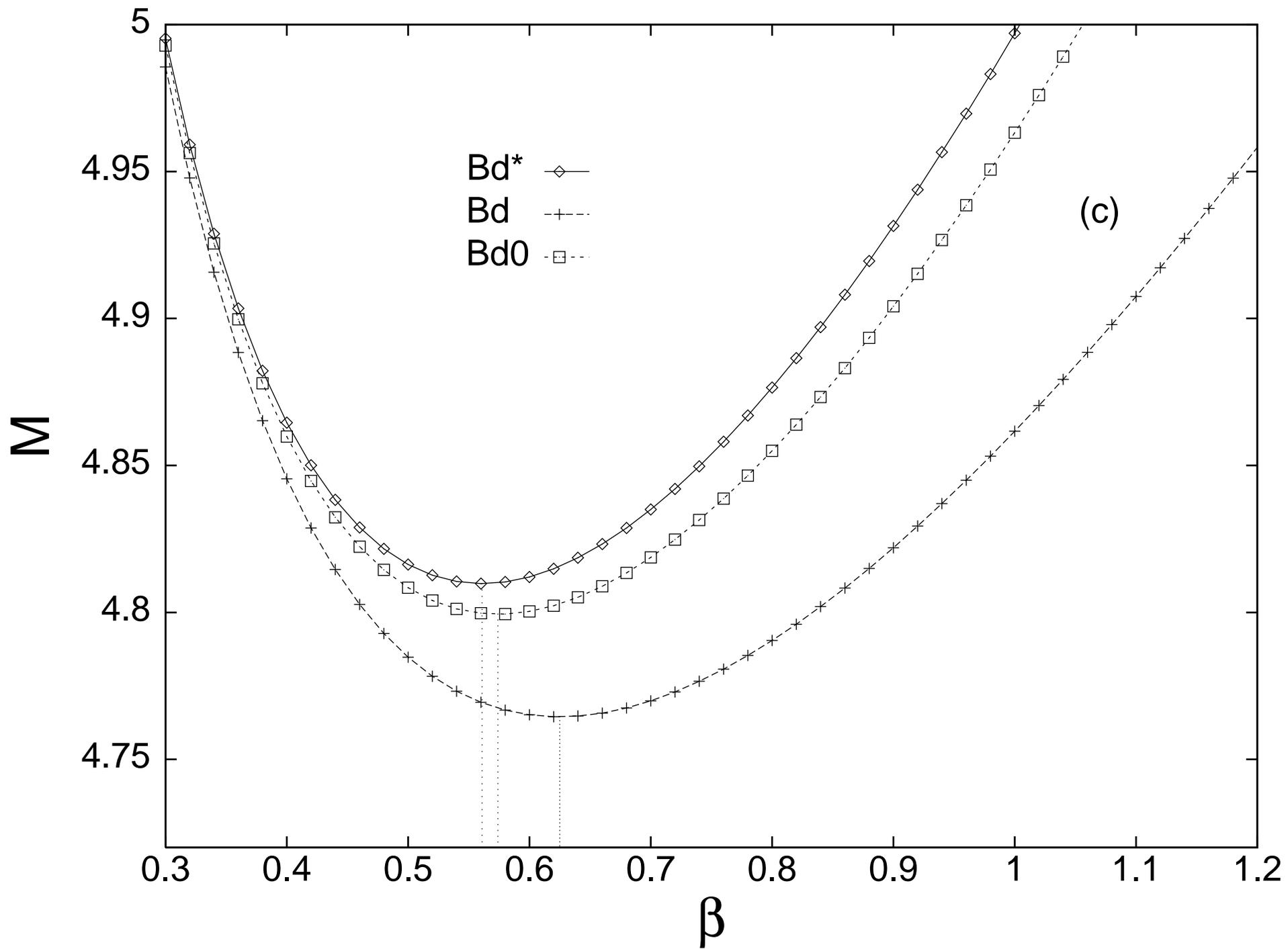
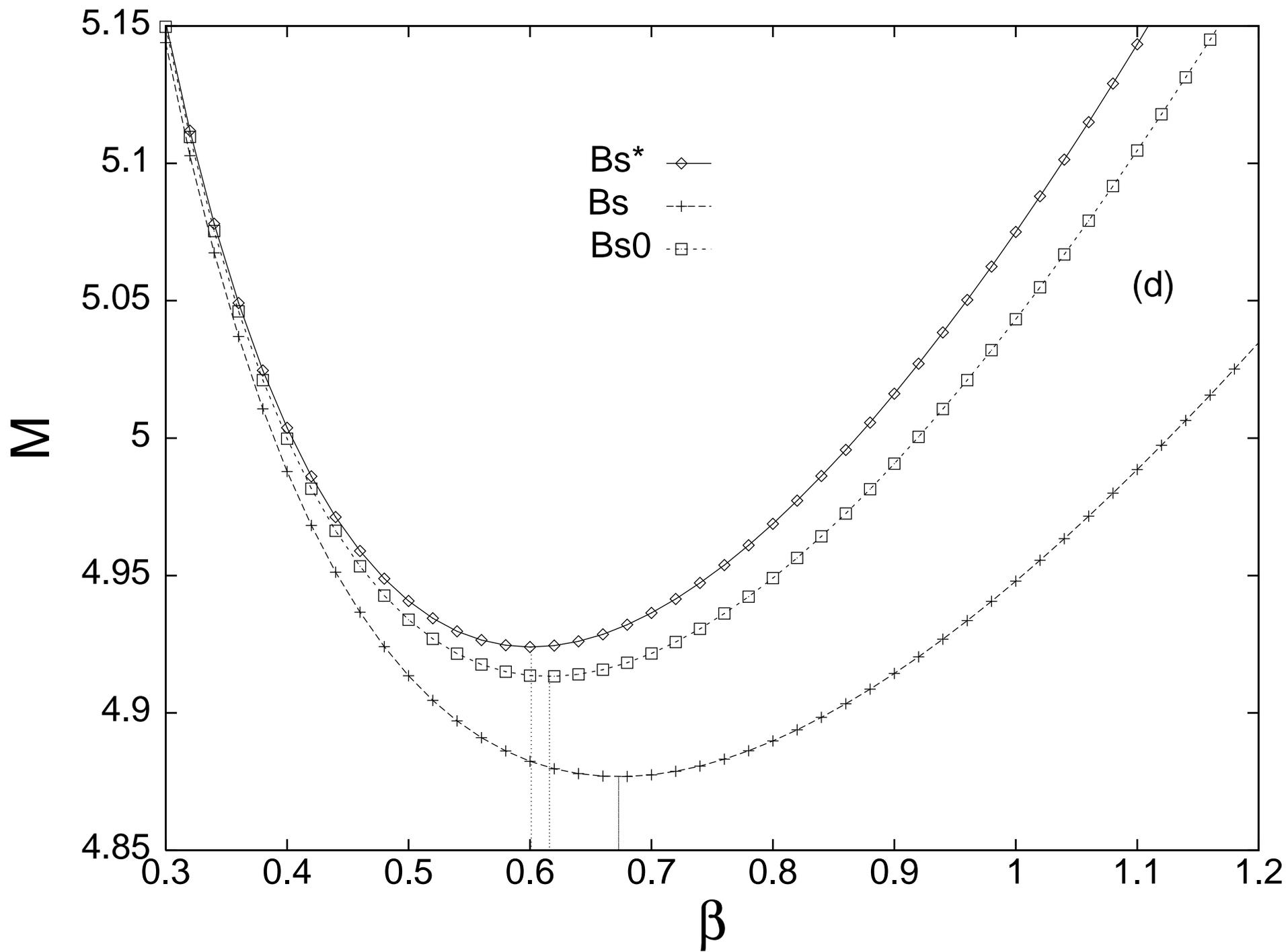
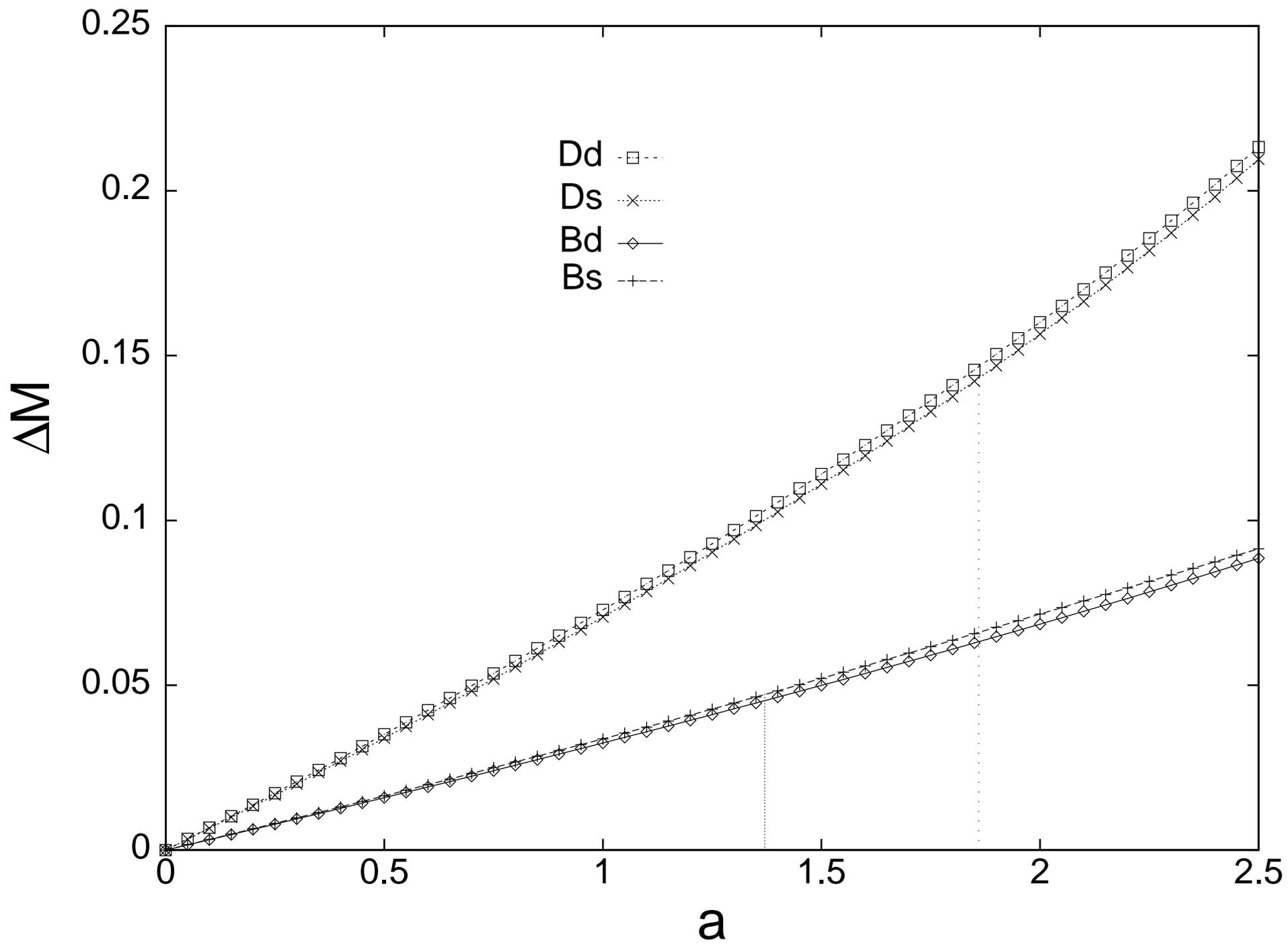
\end{document}